\documentclass[twocolumn,showkeys]{revtex4}
\usepackage[utf8]{inputenc}
\usepackage{graphicx}

\begin{document}

\title{A hydrodynamic model of Alfvenic waves and fast magneto-sound in the relativistically hot plasmas at propagation parallel to the magnetic field}

\author{Pavel A. Andreev}
\email{andreevpa@physics.msu.ru}
\affiliation{Department of General Physics, Faculty of physics, Lomonosov Moscow State University, Moscow, Russian Federation, 119991.}
\affiliation{Peoples Friendship University of Russia (RUDN University), 6 Miklukho-Maklaya Street, Moscow, 117198, Russian Federation}

\date{\today}

\begin{abstract}
The Alfvenic waves and fast magneto-sound in the relativistically hot plasmas are considered.
This study is based on the novel hydrodynamic model of plasmas with the relativistic temperatures
consisted of four equations for the material fields: the concentration and the velocity field
\emph{and} the average reverse relativistic $\gamma$ functor and the flux of the reverse relativistic $\gamma$ functor.
Three temperature regimes giving considerably different behavior of the spectrum of Alfvenic waves are found.
Four temperature regimes are obtained for the fast magneto-sound.
All these regimes show considerable variations from the nonrelativistic regime.
Dispersion equations for the electromagnetic transverse wave propagating parallel to the external magnetic field are found analytically.
Their small frequency solutions are analyzed numerically.
\end{abstract}


\maketitle





\section{Introduction}

The low frequency transverse electromagnetic waves (the Alfvenic waves and fast magneto-sound)
play essential role in many processes in laboratory plasmas and space plasmas
\cite{Turnbull PoF B 93}, \cite{Mahajan PoF 83}, \cite{Goertz PSS 84}, \cite{Louarn GRL 94},
\cite{Volwerk JGR 96}, \cite{Mottez JGR 11}, \cite{Genot PCE 01}, \cite{Mottez JPP 14},
\cite{Chaston GRL 99}, \cite{Mahajan PF 82},
\cite{Hasegawa PRL 75}, \cite{Bale PRL 05}, \cite{Sahraooui PRL 09}, \cite{Gekelman PoP 11}, \cite{Gekelman RSI 19}, \cite{Dorfman PoP 15}.
These fundamental waves are important for the low temperature plasmas and the relativistically hot plasmas.
There are the hydrodynamic approaches to the relativistically hot plasmas
\cite{Hazeltine APJ 2002}, \cite{Mahajan PoP 2002}, \cite{Shatashvili ASS 97}, \cite{Shatashvili PoP 99}, \cite{Shatashvili PoP 20}.
However, we apply the recently suggested hydrodynamic model \cite{Andreev 2021 05},
which is also applied to other wave phenomena \cite{Andreev 2021 06}, \cite{Andreev 2021 07}.

The model presented below can be derived by various technics.
But, we present basic definitions appearing at the direct derivation of hydrodynamics from the microscopic mechanic motion in accordance with Ref. \cite{Andreev 2021 05},
where this set of equations is suggested for the description of the relativistic plasmas.


We use an explicit form of operator giving the transition from the microscopic scale to the macroscopic scale.
We illustrate it with the presentation of the concentration of particles
$n(\textbf{r},t)$ in the arbitrary inertial frame \cite{Andreev 2021 05}
\begin{equation}\label{RHD2021ClLM concentration definition} n(\textbf{r},t)=\frac{1}{\Delta}\int_{\Delta}d\mbox{\boldmath $\xi$}\sum_{i=1}^{N}\delta(\textbf{r}+\mbox{\boldmath $\xi$}-\textbf{r}_{i}(t)). \end{equation}
This definition is suggested in Refs. \cite{Kuz'menkov 91}, \cite{Drofa TMP 96}, \cite{Andreev PIERS 2012}.
Nonrelativistic plasmas are considered in Refs. \cite{Drofa TMP 96}, \cite{Andreev PIERS 2012}.
More general concept as a background of the relativistic kinetics is given in Ref. \cite{Kuz'menkov 91}.
Presented description is a generalization of the Klimontovich method \cite{Weinberg Gr 72},
where the method of transition to the macroscopic level is presented via unspecified averaging.

Following the evolution of concentration $n(\textbf{r},t)$ (\ref{RHD2021ClLM concentration definition})
we obtain the continuity equation together with the microscopic definition of the current of particles,
which is traditionally represented via the velocity field.
Next step is the derivation of the current of particles.
This equation gives additional functions together with their microscopic definitions.
Kinematic part of the evolution of the current of particles gives the flux of the current of particles.
The dynamical part presented within the electromagnetic field.
It is considered in the mean-field (the self-consistent field) approximation.
Hence, three additional functions appear.
They are the average reverse relativistic $\gamma$ functor,
the flux of the reverse relativistic $\gamma$ functor,
and the second rank tensor of the current of the flux of the reverse relativistic $\gamma$ functor.
Further extension of the set of hydrodynamic equations includes the equations
for evolution of the average reverse relativistic $\gamma$ functor and
the flux of the reverse relativistic $\gamma$ functor.
The higher rank tensors are represented via the functions of smaller tensor rank using equations of state \cite{Andreev 2021 05}.

This paper is organized as follows.
In Sec. II the relativistic hydrodynamic equations are demonstrated.
In Sec. III the spectra of Alfvenic waves and fast magneto-sound are obtained and analyzed.
In Sec. IV a brief summary of obtained results is presented.


\section{Relativistic hydrodynamic model}

The hydrodynamic equations
for the relativistic plasmas with the relativistic temperature \cite{Andreev 2021 05}
are applied here to study the spectrum of Alfvenic waves and fast magneto-sound.
The relativistic hydrodynamic model consists of four equations (in three-vector notations).
First equation is the continuity equation
\begin{equation}\label{RHD2021ClLM cont via v} \partial_{t}n_{s}+\nabla\cdot(n_{s}\textbf{v}_{s})=0,\end{equation}
where $s=e,i$ is the index showing the species of electrons and ions.
The continuity equation does not contain any contribution of interaction.
The dynamics of concentration $n$ depend on the evolution of the velocity field.
So, we need to present the equation for the velocity field evolving with time.
Let us mention that the velocity field is proportional to the current of particles.
It has no simple relation with the momentum density.
Following Ref. \cite{Andreev 2021 05} we present the velocity field evolution equation:
$$m_{s}n_{s}[\partial_{t}v^{a}_{s}+(\textbf{v}_{s}\cdot\nabla)v^{a}_{s}]+\partial^{a}p_{s}
=q_{s}\Gamma_{s} E^{a}$$
\begin{equation}\label{RHD2021ClLM Euler for v}
+\frac{q_{s}}{c}\varepsilon^{abc}(\Gamma_{s} v_{s}^{b}+t_{s}^{b})B^{c}
-\frac{q_{s}}{c^{2}}[(\Gamma_{s} v_{s}^{a} v_{s}^{b}+v_{s}^{a}t_{s}^{b}+v_{s}^{b}t_{s}^{a})E^{b}
+\tilde{t}_{s}E^{a}], \end{equation}
where $p_{s}$ is the flux of the thermal velocities.
The interaction is found in the mean-field approximation.
The equation of evolution of the averaged reverse relativistic gamma factor, called here the hydrodynamic Gamma function, is
\begin{equation}\label{RHD2021ClLM eq for Gamma} \partial_{t}\Gamma_{s}+\partial_{b}(\Gamma_{s} v_{s}^{b}+t_{s}^{b})
=-\frac{q_{s}n_{s}}{m_{s}c^{2}}\textbf{v}_{s}\cdot\textbf{E}\biggl(1-\frac{1}{c^{2}}\biggl(\textbf{v}_{s}^{2}+\frac{5p_{s}}{n_{s}}\biggr)\biggr).\end{equation}
The averaged reverse relativistic gamma factor appears on the right-hand side of the velocity field evolution equation.
It appears along with the vector of current of the reverse relativistic gamma factor
and the second rank tensor describing the flux of current of the reverse relativistic gamma factor.

The fourth equation in the presented set of hydrodynamic equations is
the equation for the thermal part of the
current of the reverse relativistic gamma factor
(thermal part of the hydrodynamic Theta function):
$$(\partial_{t}+\textbf{v}_{s}\cdot\nabla)t_{s}^{a}+\partial_{a}\tilde{t}_{s}
+(\textbf{t}_{s}\cdot\nabla) v_{s}^{a}+t_{s}^{a} (\nabla\cdot \textbf{v}_{s})$$
$$+\Gamma_{s}(\partial_{t}+\textbf{v}_{s}\cdot\nabla)v^{a}_{s}
=\frac{q_{s}}{m}n_{s}E^{a}\biggl[1-\frac{\textbf{v}_{s}^{2}}{c^{2}}-\frac{3p_{s}}{n_{s}c^{2}}\biggr]$$
$$+\frac{q_{s}}{m_{s}c}\varepsilon^{abc}n_{s}v_{s}^{b}B^{c}\biggl[1-\frac{\textbf{v}_{s}^{2}}{c^{2}}-\frac{5p_{s}}{n_{s}c^{2}}\biggr]
-\frac{2q_{s}}{m_{s}c^{2}}E^{a}p_{s}\biggl[1-\frac{\textbf{v}_{s}^{2}}{c^{2}}\biggr]$$
\begin{equation}\label{RHD2021ClLM eq for t a} -\frac{q_{s}}{m_{s}c^{2}}n_{s}v_{s}^{a}v_{s}^{b}E^{b}\biggl[1-\frac{\textbf{v}_{s}^{2}}{c^{2}}-\frac{9p_{s}}{n_{s}c^{2}}\biggr]
-\frac{10q_{s}}{3m_{s}c^{4}}M_{s} E^{a}.\end{equation}

The equations for the material fields (\ref{RHD2021ClLM cont via v})-(\ref{RHD2021ClLM eq for t a})
are coupled to the Maxwell equations
$ \nabla \cdot\textbf{B}=0$, $\nabla\times \textbf{E}=-\frac{1}{c}\partial_{t}\textbf{B}$,
\begin{equation}\label{RHD2021ClLM rot E and div E}
\nabla \cdot\textbf{E}=4\pi(q_{e}n_{e}+q_{i}n_{i}),
\end{equation}
and
\begin{equation}\label{RHD2021ClLM rot B with time}
\nabla\times \textbf{B}=\frac{1}{c}\partial_{t}\textbf{E}
+\frac{4\pi q_{e}}{c}n_{e}\textbf{v}_{e}
+\frac{4\pi q_{i}}{c}n_{i}\textbf{v}_{i}.\end{equation}


\begin{figure}[h!]
\includegraphics[width=8cm,angle=0]{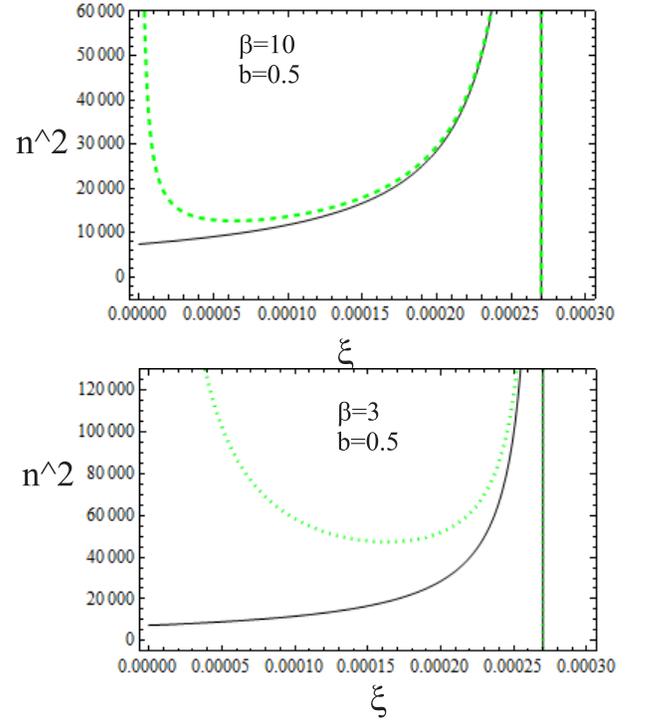}
\caption{\label{RHD2021ClLM Fig 01}
The square of the refractive index of the Alfvenic waves is demonstrated.
Each figure contains the square of the refractive index in relativistic regime with the temperature characterized by parameter $\beta=m_{e}c^{2}/T$,
which is plotted within the dashed green line.
Plasmas under consideration are isothermal $T_{e}=T_{i}=T$.
Temperature is given in the energy units.
It is compared with the nonrelativistic regime presented within the black continuous line.
Here we have two regimes corresponding to the relativistically hot electrons $T\sim m_{e}c^{2}$: $\beta=10$ and $\beta=3$.
Due to the large mass of ions it corresponds to the nonrelativistic regime for ions.
Large value of the parameter $\beta$ (on the upper figure we have $\beta=10$) corresponds to small relativistic effects for electrons.}
\end{figure}

\begin{figure}[h!]
\includegraphics[width=8cm,angle=0]{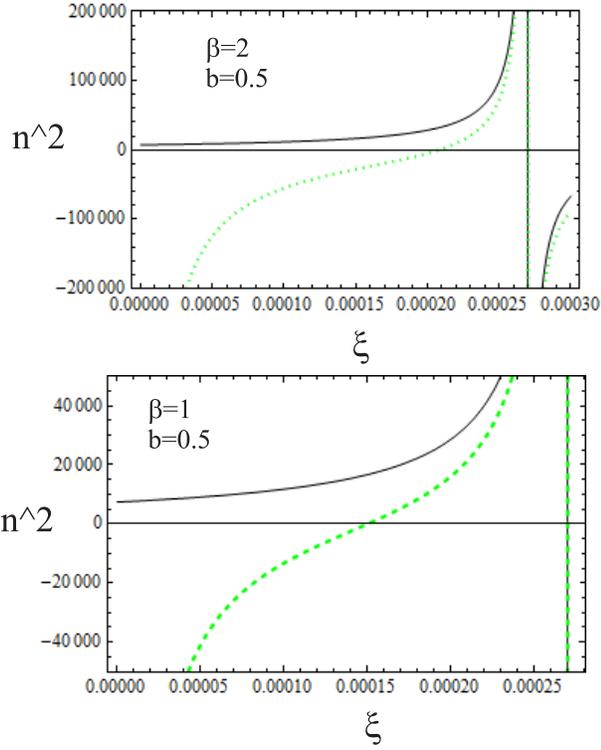}
\caption{\label{RHD2021ClLM Fig 02}
The square of the refractive index of the Alfvenic waves is demonstrated.
The further increase of temperature in compare with Fig. 1 is presented.
Two temperatures are used for two plots in this figure $\beta=2$ and $\beta=1$. }
\end{figure}

\section{Alfvenic waves and fast magneto-sound in the relativistically hot plasmas}

High-frequency regime for the electromagnetic transverse wave propagating parallel to the external magnetic field is considered in Ref. \cite{Andreev 2021 06},
where the motion of ions is neglected.
Here we consider the small-frequency excitations (Alfvenic waves and fast magneto-sound) in the relativistically hot plasmas
and focus our attention on the contribution of ions.

We consider small amplitude collective excitations of the macroscopically motionless equilibrium state.
This equilibrium state is described by the relativistic Maxwellian distribution.
The equilibrium state is described within equilibrium concentrations $n_{0e}=n_{0i}\equiv n_{0}$.
The velocity fields $\textbf{v}_{0e,i}$ in the equilibrium state are equal to zero.
The equilibrium electric field $\textbf{E}_{0}$ is also equal to zero.
The plasma is located in the constant and uniform external magnetic field $\textbf{B}_{0}=B_{0}\textbf{e}_{z}$.
Two second rank tensors and one fourth rank tensor are involved in the description of the thermal effects.
The symmetric tensors $p_{e,i}^{ab}$ and $t_{e,i}^{ab}$ are assumed to be diagonal tensors:
$p_{e,i}^{ab}=p_{e,i}\delta^{ab}$ and $t_{e,i}^{ab}=\tilde{t}_{e,i}\delta^{ab}$.
The "diagonal" form is assumed for the symmetric fourth rank tensors $M_{e,i}^{abcd}$ as well:
$M_{e,i}^{abcd}=M_{0e,i}(\delta^{ab}\delta^{cd}+\delta^{ac}\delta^{bd}+\delta^{ad}\delta^{bc})/3$.

The following equilibrium functions
$\Gamma_{0e,i}$, $\textbf{t}_{0e,i}$, $p_{0e,i}$, $\tilde{t}_{0e,i}$, $\textbf{q}_{0e,i}$, $M_{0e,i}$
are involved in the description of the equilibrium state
$p_{e,i}^{ab}=c^{2}\delta^{ab}\tilde{Z}_{e,i}f_{1e,i}(\beta)/3$,
$t_{e,i}^{ab}=c^{2}\delta^{ab}\tilde{Z}_{e,i}f_{2e,i}(\beta)/3$
$M^{abcd}_{e,i}=c^{4}(\delta^{ab}\delta^{cd}+\delta^{ac}\delta^{bd}+\delta^{ad}\delta^{bc})\tilde{Z}_{e,i}f_{3e,i}(\beta)/15$,
$\Gamma_{0e,i}=n_{0}K_{1}(\chi_{e,i}\beta)/K_{2}(\chi_{e,i}\beta)$,
$\chi_{e}=1$,
$\chi_{i}=(m_{i}/m_{e})\cdot(T_{e}/T_{i})$,
and
$\textbf{q}_{e,i}=0$,
where
$\beta=m_{e}c^{2}/T_{e}$ is the reverse dimensionless temperature of electrons,
$\tilde{Z}_{e,i}=4\pi Z_{e,i} (m_{e,i}c)^{3}=n_{e,i}\chi_{e,i}\beta K_{2}^{-1}(\chi_{e,i}\beta)$,
\begin{equation}\label{RHD2021ClLM f 1} f_{1e,i}(\beta)=f_{1}(\chi_{e,i}\beta)=\int_{1}^{+\infty}\frac{d x}{x}(x^{2}-1)^{3/2} e^{-\chi_{e,i}\beta x}, \end{equation}
\begin{equation}\label{RHD2021ClLM f 2} f_{2e,i}(\beta)=f_{2}(\chi_{e,i}\beta)=\int_{1}^{+\infty}\frac{d x}{x^{2}}(x^{2}-1)^{3/2} e^{-\chi_{e,i}\beta x}, \end{equation}
and
\begin{equation}\label{RHD2021ClLM f 3} f_{3e,i}(\beta)=f_{3}(\chi_{e,i}\beta)=\int_{1}^{+\infty}\frac{d x}{x^{3}}(x^{2}-1)^{5/2} e^{-\chi_{e,i}\beta x}. \end{equation}
Factor $\chi_{i}$ shows representation of the dimensionless temperature of ions (in units of $m_{i}c^{2}$)
via the dimensionless temperature of electrons (in units of $m_{e}c^{2}$).
Functions $f_{1}(\beta)$, $f_{2}(\beta)$ and $f_{3}(\beta)$ are introduced in Ref. \cite{Andreev 2021 05} for the electrons.
Hence, dimensionless parameter $x$ is the energy of electrons $\varepsilon_{e}$ over the rest energy of electron $x=\varepsilon_{e}/m_{e}c^{2}$.
Same functions for ions have different definition of the dimensionless parameter under integrals,
where $x=\varepsilon_{i}/m_{i}c^{2}$.
So, the exponent indicator in the distribution function includes additional factor
$\varepsilon_{i}/T_{i}=(\varepsilon_{i}/m_{i}c^{2})(T_{i}/m_{i}c^{2})^(-1)$$=(\varepsilon_{i}/m_{i}c^{2})(T_{i}/m_{i}c^{2})^(-1)$.
Functions $f_{1}(\chi_{e,i}\beta)$, $f_{2}(\chi_{e,i}\beta)$ and $f_{3}(\chi_{e,i}\beta)$ are calculated numerically below for the chosen values of temperatures.
We introduce three characteristic velocities
$\delta p_{e,i}=U_{pe,}^{2} \delta n_{e,i}$,
$\delta \tilde{t}_{e,i}=U_{te,i}^{2} \delta n_{e,i}$,
and
$\delta M_{e,i}=U_{Me,i}^{4} \delta n_{e,i}$.

\begin{figure}[h!]
\includegraphics[width=8cm,angle=0]{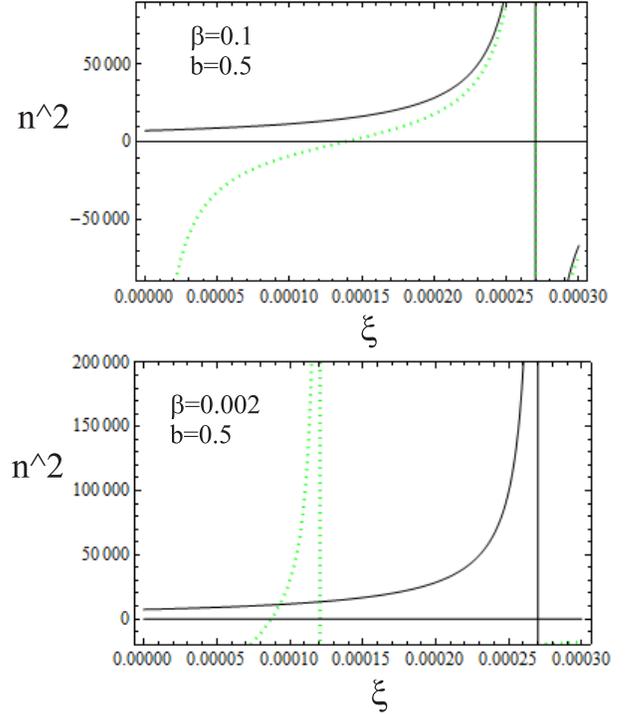}
\caption{\label{RHD2021ClLM Fig 03}
The square of the refractive index of the Alfvenic waves is demonstrated.
The further increase of temperature in compare with Fig. 2 is presented.
Two temperatures are used for two plots in this figure $\beta=0.1$ and $\beta=0.002$.}
\end{figure}

\begin{figure}[h!]
\includegraphics[width=8cm,angle=0]{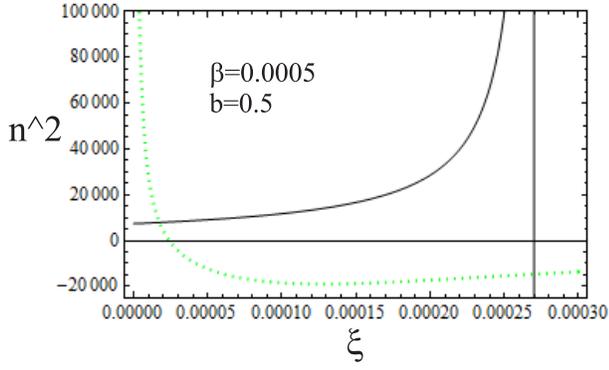}
\caption{\label{RHD2021ClLM Fig 04}
The square of the refractive index of the Alfvenic waves is demonstrated.
The further increase of temperature in compare with Fig. 3 is presented.
One temperatures is used for this figure $\beta=0.0005$.}
\end{figure}

The linearized equations have same structure for all species.
Hence, equations for electrons and ions repeat equations presented for electrons in Ref. \cite{Andreev 2021 06}.

Let us present the expressions for the velocity field entering the linearized Maxwell equations
$$v_{xe,i}=\frac{\imath q_{e,i}}{m_{e,i}}\times$$
\begin{equation}\label{RHD2021ClLM v x a}
\frac{\omega(\frac{\Gamma_{0e,i}}{n_{0e,i}}-\frac{U_{te,i}^{2}}{c^{2}})\delta E_{x}
-\imath\Omega_{e,i}\delta E_{y}(1-\frac{5U_{pe,i}^{2}}{c^{2}}-\frac{10U_{Me,i}^{4}}{3c^{4}})}{\omega^{2}-\Omega_{e,i}^{2}(1-5U_{pe,i}^{2}/c^{2})}, \end{equation}
and
$$v_{ye,i}=\frac{\imath q_{e,i}}{m_{e,i}}\times$$
\begin{equation}\label{RHD2021ClLM v y a}
\frac{\imath\Omega_{e,i}(1-\frac{5U_{pe,i}^{2}}{c^{2}}-\frac{10U_{Me,i}^{4}}{3c^{4}})\delta E_{x}
+\omega(\frac{\Gamma_{0e,i}}{n_{0e,i}}-\frac{U_{te,i}^{2}}{c^{2}})\delta E_{y}}{\omega^{2}-\Omega_{e,i}^{2}(1-5U_{pe,i}^{2}/c^{2})}, \end{equation}
where
$\Omega_{e}=-\mid \Omega_{e}\mid=q_{e}B_{0}/m_{e}c$,
$\Omega_{i}=\mid \Omega_{i}\mid=q_{i}B_{0}/m_{i}c$
are the electron and ion cyclotron frequencies.

The linearized Maxwell equations are
\begin{equation}\label{RHD2021ClLM Maxwell lin wave}
(\omega^{2}-k_{z}^{2}c^{2})\delta \textbf{E}+4\pi\imath\omega \sum_{s=e,i}q_{s}n_{0s}\delta \textbf{v}_{s}=0. \end{equation}

Using expressions of the velocity field via the electric field in equation (\ref{RHD2021ClLM Maxwell lin wave})
we obtain the equations for the projections of the electric field $\hat{\varepsilon}\delta \textbf{E}=0$:
\begin{equation}\label{RHD2021ClLM Dispersion determinant}
\hat{\varepsilon}=
\left(
  \begin{array}{cc}
    \varepsilon_{xx}, & \varepsilon_{xy} \\
    \varepsilon_{yx}, & \varepsilon_{yy} \\
  \end{array}
\right),
\end{equation}
where
$$\varepsilon_{xx}=\varepsilon_{yy}=
-\omega^{2}+k^{2}c^{2}$$
$$+\frac{\omega^{2}\cdot \omega^{2}_{Le}}{\omega^{2}-\Omega^{2}_{e}(1-5U_{pe}^{2}/c^{2})}
\biggl(\frac{\Gamma_{0e}}{n_{0e}}-\frac{U_{te}^{2}}{c^{2}}\biggr)$$
\begin{equation}\label{RHD2021ClLM epsilon xx}
+\frac{\omega^{2}\cdot \omega^{2}_{Li}}{\omega^{2}-\Omega^{2}_{i}(1-5U_{pi}^{2}/c^{2})}
\biggl(\frac{\Gamma_{0i}}{n_{0i}}-\frac{U_{ti}^{2}}{c^{2}}\biggr)
\end{equation}
and
$$\varepsilon_{xy}=(\varepsilon_{yx})^*$$
$$=
\frac{-\imath\omega\Omega_{e}\cdot \omega^{2}_{Le}}{\omega^{2}-\Omega^{2}_{e}(1-5U_{pe}^{2}/c^{2})}
\biggl(1-\frac{5U_{pe}^{2}}{c^{2}}-\frac{10U_{Me}^{4}}{3c^{4}}\biggr)$$
\begin{equation}\label{RHD2021ClLM epsilon xy}
+\frac{-\imath\omega\Omega_{i}\cdot \omega^{2}_{Li}}{\omega^{2}-\Omega^{2}_{i}(1-5U_{pi}^{2}/c^{2})}
\biggl(1-\frac{5U_{pi}^{2}}{c^{2}}-\frac{10U_{Mi}^{4}}{3c^{4}}\biggr)
\end{equation}
where
$\omega_{Le}^{2}=4\pi e^{2}n_{0}/m_{e}$ is the electron Langmuir frequency,
$\omega_{Li}^{2}=4\pi e^{2}n_{0}/m_{i}$ is the ion Langmuir frequency,
and symbol $()^*$ shows the complex conjugation.



\subsection{Nonrelativistic limit}

Let us prepare background for the analysis of the relativistically hot plasmas
by consideration of the limit of small nonrelativistic temperatures.
To this end, we neglect the thermal velocities $U_{p}$, $U_{t}$, $U_{M}$ in compare with the speed of light in equations
(\ref{RHD2021ClLM Dispersion determinant}), (\ref{RHD2021ClLM epsilon xx}), and (\ref{RHD2021ClLM epsilon xy}).
Hence, we obtain
$$\hat{\varepsilon}_{NR}=$$
\begin{equation}\label{RHD2021ClLM Dispersion determinant nonRel}
\left(
  \begin{array}{cc}
k^{2}c^{2}-\omega^{2} +\sum_{s}\frac{\omega^{2}\cdot \omega^{2}_{Ls}}{\omega^{2}-\Omega^{2}_{s}},
    & \sum_{s}\frac{-\imath\Omega_{s}\omega\cdot \omega^{2}_{Ls}}{\omega^{2}-\Omega^{2}_{s}}\\
    \sum_{s}\frac{\imath\Omega_{s}\omega\cdot \omega^{2}_{Ls}}{\omega^{2}-\Omega^{2}_{s}},
    &k^{2}c^{2}-\omega^{2} +\sum_{s}\frac{\omega^{2}\cdot \omega^{2}_{Ls}}{\omega^{2}-\Omega^{2}_{s}}\\
  \end{array}
\right),
\end{equation}
where $s=e,i$.

The dielectric permeability tensor (\ref{RHD2021ClLM Dispersion determinant nonRel}) gives two expressions for the refractive index.
We present one of them
which includes the Alfvenic wave
\begin{equation}\label{RHD2021ClLM Spectrum}
n^{2}=1
-\frac{1}{\omega}\frac{\omega_{Le}^{2}}{\omega +\mid\Omega_{e}\mid}
-\frac{1}{\omega}\frac{\omega_{Li}^{2}}{\omega -\Omega_{i}}.
\end{equation}

The spectrum of nonrelativistic Alfvenic wave is located in area $\omega\in(0,\Omega_{i})$.
However, if we consider the limit $\omega\ll\Omega_{i}$
we get the famous expression
\begin{equation}\label{RHD2021ClLM Alfvenic wave spectrum} \omega^{2}=\frac{k^{2}v_{A}^{2}}{1+v_{A}^{2}/c^{2}},\end{equation}
where $v_{A}\equiv c\Omega_{i}/\omega_{Li}$ is the Alfven velocity.

Expression (\ref{RHD2021ClLM Alfvenic wave spectrum}) is obtained from (\ref{RHD2021ClLM Dispersion determinant nonRel})
in the low frequency regime
$$\hat{\varepsilon}_{NR}=$$
\begin{equation}\label{RHD2021ClLM Dispersion determinant nonRel low freq}
\left(
  \begin{array}{cc}
k^{2}c^{2}-\omega^{2} -\sum_{s}\frac{\omega^{2}\cdot \omega^{2}_{Ls}}{\Omega^{2}_{s}},
    & \sum_{s}\frac{\imath\omega\cdot \omega^{2}_{Ls}}{\Omega_{s}}\\
    \sum_{s}\frac{-\imath\omega\cdot \omega^{2}_{Ls}}{\Omega_{s}},
    &k^{2}c^{2}-\omega^{2} -\sum_{s}\frac{\omega^{2}\cdot \omega^{2}_{Ls}}{\Omega^{2}_{s}}\\
  \end{array}
\right),
\end{equation}
where
$\varepsilon_{NR,xy}=\sum_{s}\frac{\imath\omega\cdot \omega^{2}_{Ls}}{\Omega_{s}}=0$,
and
$\frac{\omega^{2}_{Le}}{\Omega^{2}_{e}}\ll \frac{\omega^{2}_{Li}}{\Omega^{2}_{i}}$
in $\varepsilon_{NR,xx}$ and $\varepsilon_{NR,yy}$.

\begin{figure}[h!]
\includegraphics[width=8cm,angle=0]{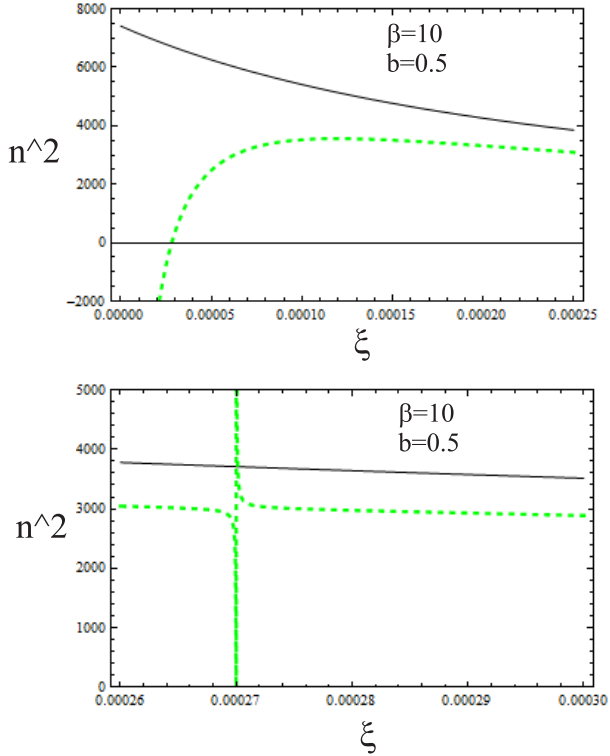}
\caption{\label{RHD2021ClLM Fig 05}
The refractive index for the fast magneto-sound is presented.
The continuous line presents the nonrelativistic result.
The result of the relativistic model is presented with the dashed line.
These notations are saved in other figures below.
The reverse dimensionless temperature is shown in the plot.
}
\end{figure}
\begin{figure}[h!]
\includegraphics[width=8cm,angle=0]{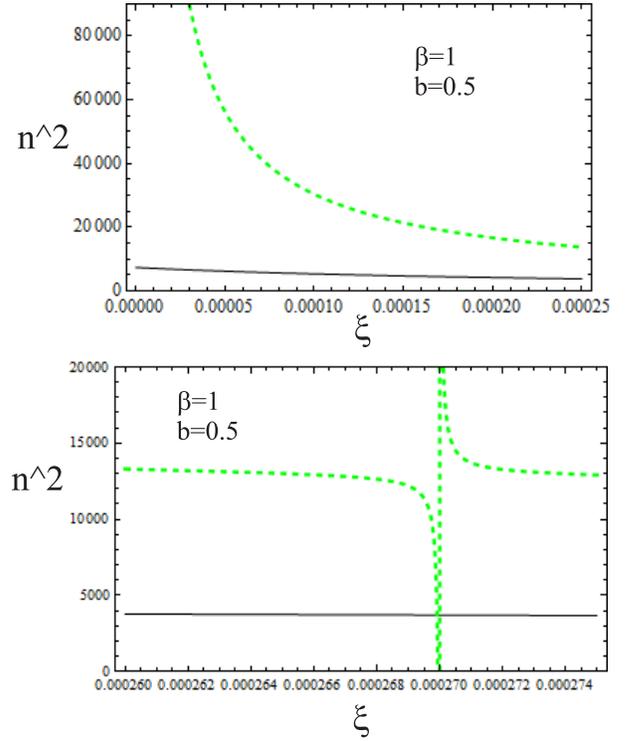}
\caption{\label{RHD2021ClLM Fig 06}
The refractive index for the fast magneto-sound is given.
More accurately, we have the curves for the low-frequency part of the fast extraordinary wave
and the small frequency second branch.
The continuous line presents the nonrelativistic result.
The result of the relativistic model is presented with the dashed line.
}
\end{figure}
\begin{figure}[h!]
\includegraphics[width=8cm,angle=0]{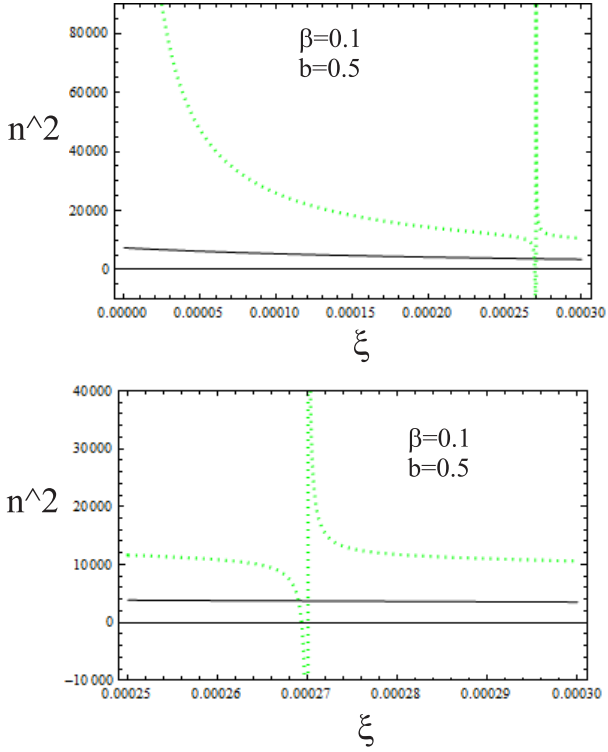}
\caption{\label{RHD2021ClLM Fig 07}
The refractive index for the fast magneto-sound is obtained
(the low-frequency part of the fast extraordinary wave
and the small frequency second branch like on Fig. 6).
The continuous line presents the nonrelativistic result.
The result of the relativistic model is presented with the dashed line.
}
\end{figure}

\begin{figure}[h!]
\includegraphics[width=8cm,angle=0]{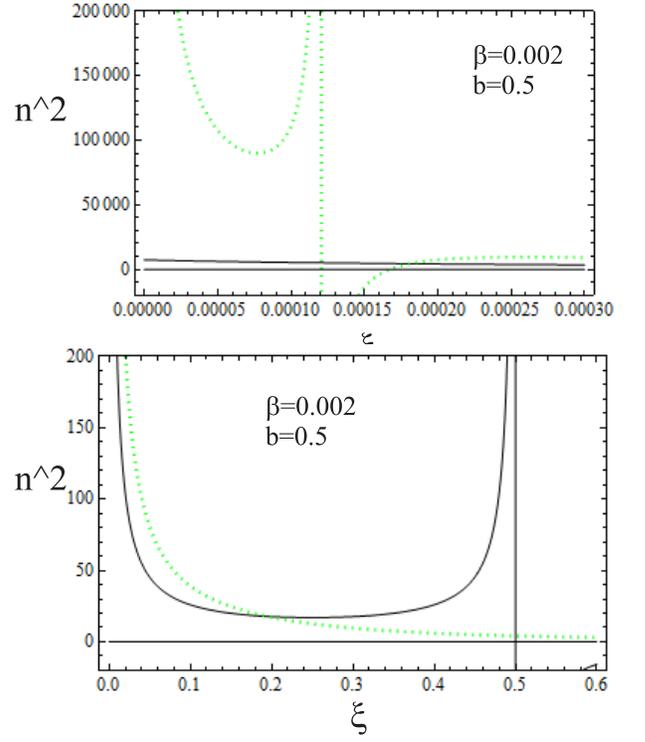}
\caption{\label{RHD2021ClLM Fig 08}
The refractive index for the fast magneto-sound is presented
which is actually replaced by
the low-frequency part of the fast extraordinary wave
and the small frequency second branch.
The continuous line presents the nonrelativistic result.
The result of the relativistic model is presented with the dashed line.
}
\end{figure}

\begin{figure}[h!]
\includegraphics[width=8cm,angle=0]{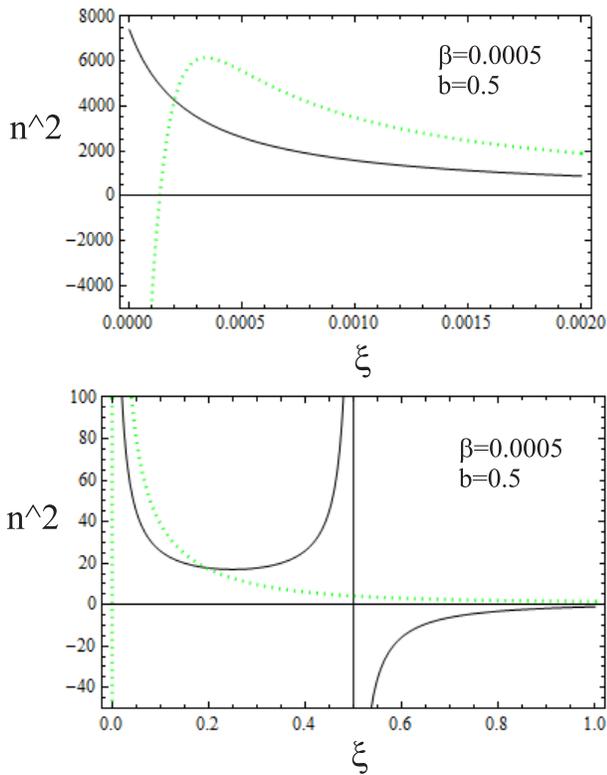}
\caption{\label{RHD2021ClLM Fig 09}
The refractive index for the fast magneto-sound is shown,
but in this regime it is completely replaced
by the low-frequency part of the fast extraordinary wave.
The continuous line presents the nonrelativistic result.
The result of the relativistic model is presented with the dashed line.}
\end{figure}

\subsection{Relativistic Alfvenic waves}

Let us get back to the relativistic temperature regime.
The refractive index for the Alfvenic waves following from equations
(\ref{RHD2021ClLM Dispersion determinant}), (\ref{RHD2021ClLM epsilon xx}), (\ref{RHD2021ClLM epsilon xy})
has the form presented below
$$n^{2}=1-\frac{1}{\omega}\frac{\omega_{Le}^{2}}{\omega^{2}-\Omega_{e}^{2}(1-5\frac{U_{pe}^{2}}{c^{2}})}\times$$
$$\times\biggl[\omega\biggl(\frac{\Gamma_{0e}}{n_{0e}}-\frac{U_{te}^{2}}{c^{2}}\biggr) -\mid\Omega_{e}\mid\biggl(1-5\frac{U_{pe}^{2}}{c^{2}}-\frac{10U_{Me}^{4}}{3c^{4}}\biggr)\biggr]$$
$$-\frac{1}{\omega}\frac{\omega_{Li}^{2}}{\omega^{2}-\Omega_{i}^{2}(1-5\frac{U_{pi}^{2}}{c^{2}})}\times$$
\begin{equation}\label{RHD2021ClLM Spectrum AlW}
\times\biggl[\omega\biggl(\frac{\Gamma_{0i}}{n_{0i}}-\frac{U_{ti}^{2}}{c^{2}}\biggr) +\Omega_{i}\biggl(1-5\frac{U_{pi}^{2}}{c^{2}}-\frac{10U_{Mi}^{4}}{3c^{4}}\biggr)\biggr].
\end{equation}
Equation (\ref{RHD2021ClLM Spectrum AlW}) is a generalization of equation (\ref{RHD2021ClLM Spectrum}).

High frequency part of the refractive index $n(\omega)$,
which is located near the ion-cyclotron frequency (the vertical line),
shows small variation in compare with the nonrelativistic regime.
Here, the relativistic effects show small increase of the refractive index.
However, the transition to the small frequency are demonstrates the considerable increase of frequency.
Moreover, the zero frequency limit is completely different.
Instead of monotonic decrease of $n$ at the decrease of frequency $\omega$
we see asymptotic growth to the infinity.
It is related to the nonzero value of the nondiagonal elements of the dielectric permeability tensor.
So, the pole at $\omega=0$ is not canceled in this regime in contrast with the nonrelativistic limit.

Small increase of temperature from $\beta=3$ to $\beta=2$ and $\beta=1$
shows a qualitative change of $n(\omega)$ in full area of existence of the Alfvenic wave.
The increase of the refractive index demonstrated above changes to its considerable decrease.
It happens due to the fact that $n(\omega)$ goes to zero at $\omega_{0}\in (0,\Omega_{i})$.
The pole of $n(\omega)$ at $\omega=0$ remains,
but signature of $n^{2}(\omega)$ changes.
All of it reveals in the decrease of the area of existence of the Alfvenic wave
from $\omega\in (0,\Omega_{i})$ to $\omega\in (\omega_{0},\Omega_{i})$.
Same result can be found at the further increase of temperature $\beta=0.1$ (see the upper plot in Fig. 3),
where electrons getting in the ultrarelativistic regime,
but ions are still in the nonrelativistic regime.

Parameter $\beta=0.002$ corresponds to the ultrarelativistic electrons and weakly relativistic ions (see the lower figure in Fig. 3).
Relativistic temperature of ions reveals in the difference of the effective cyclotron frequency
$\Omega_{i}\sqrt{(1-\frac{5U_{pe,i}^{2}}{c^{2}})}$ from the bare ion-cyclotron frequency $\Omega_{i}$.
Hence, the addition restriction of area of the Alfvenic wave existence appears.
However, general picture of $n^{2}(\omega)$ is the same as at $\beta=0.1$.

Fig. 4 is plotted for the regime of relativistically hot ions.
Here, the effective cyclotron frequency square
$\Omega_{i}^{2}(1-\frac{5U_{pe,i}^{2}}{c^{2}})$ goes to small negative value.
Area of existence of the Alfvenic wave disappear.
However, there is a wave in area $\omega\in (0,\omega')$,
where $\omega'\ll\Omega_{i}$.
In its area of existence its $n(\omega)$ is similar to the small frequency part of the Alfvenic wave demonstrated in Fig. 1.

Therefore, three temperature regimes are found for the relativistic Alfven waves in addition to the low (nonrelativistic) temperature regime.

\subsection{Relativistic fast magneto-sound}

High-frequency part of fast magneto-sound can be described with no account of ions.
Corresponding result is shown in Ref. \cite{Andreev 2021 06} for the relativistically hot electrons.
Ref. \cite{Andreev 2021 06} demonstrates that
$n\rightarrow+\infty$ at $\omega\rightarrow0$ for fast magneto-sound at motionless ions.
This pole shifts up to
$\Omega_{i}\sqrt{(1-\frac{5U_{pe,i}^{2}}{c^{2}})}$
at the account of the relativistic ion motion
(let us remind that $n(\omega=0)$ has finite positive value for the nonrelativistic plasmas).

The refractive index for the relativistic fast magneto-sound including the relativistic ion motion is
$$n^{2}=1-\frac{1}{\omega}\frac{\omega_{Le}^{2}}{\omega^{2}-\Omega_{e}^{2}(1-5\frac{U_{pe}^{2}}{c^{2}})}\times$$
$$\times\biggl[\omega\biggl(\frac{\Gamma_{0e}}{n_{0e}}-\frac{U_{te}^{2}}{c^{2}}\biggr) +\mid\Omega_{e}\mid\biggl(1-5\frac{U_{pe}^{2}}{c^{2}}-\frac{10U_{Me}^{4}}{3c^{4}}\biggr)\biggr]$$
$$-\frac{1}{\omega}\frac{\omega_{Li}^{2}}{\omega^{2}-\Omega_{i}^{2}(1-5\frac{U_{pi}^{2}}{c^{2}})}\times$$
\begin{equation}\label{RHD2021ClLM Spectrum FMS}
\times\biggl[\omega\biggl(\frac{\Gamma_{0i}}{n_{0i}}-\frac{U_{ti}^{2}}{c^{2}}\biggr) -\Omega_{i}\biggl(1-5\frac{U_{pi}^{2}}{c^{2}}-\frac{10U_{Mi}^{4}}{3c^{4}}\biggr)\biggr].
\end{equation}

If the ion temperature is relatively small
we have pole at $\Omega_{i}$.
This pole splits the area of existence of the fast magneto-sound on two areas separated by small interval of frequencies near $\Omega_{i}$.
Hence, we can conclude that
there are two branches of waves instead of single fast magneto-sound.
Frequencies of one wave is located below the effective cyclotron frequency
$\Omega_{i}\sqrt{(1-\frac{5U_{pe,i}^{2}}{c^{2}})}$
(we call this wave the small
frequency second branch).
Frequency of the second wave is located above
$\Omega_{i}\sqrt{(1-\frac{5U_{pe,i}^{2}}{c^{2}})}$.
We associate it with the fast magneto-sound,
since it has same high frequency limit of the fast magneto-sound.

Fig. 5 shows the result of numerical analysis of the refractive index for the fast magneto-sound.
It demonstrates that
the small relativistic temperature of electrons at almost nonrelativistic ions
leads to the considerable change of the refractive index.
The lower figure in Fig. 5 shows area near the ion cyclotron frequency $\Omega_{i}$.
The refractive index $n(\omega)$ has a pole here in the relativistic regime.
There is small interval below the ion cyclotron frequency,
where no wave exist.
This area separate areas of existence of two waves.
The fast magneto-sound is restricted from the small frequency area by the ion cyclotron frequency.
The small frequency second branch appears in area $\omega\in (\tilde{\omega},\Omega_{i})$.
For $\beta=10$ we obtain $\tilde{\omega}/_{Le}=0.00003$ as it is shown in the upper Fig. 5.
Moreover, the refractive index in the relativistic regime is smaller
then the nonrelativistic the refractive index.

The second relativistic regime appears at $\beta=1$ (see Fig. 6) and $\beta=0.1$ (see Fig. 7).
It modifies both the fast magneto-sound and
the small frequency second branch.
The fast magneto-sound wave exist below the effective electron cyclotron frequency,
but its square is below zero.
Hence, this area disappears.
We have the low frequency part of the fast-extraordinary wave
located at the frequency above the ion cyclotron frequency.
The relativistic refractive index is larger
then the refractive index in the nonrelativistic case.
The small frequency second branch exists in area $\omega\in(0,\Omega_{i})$,
so the area is extended down to zero frequency.
It corresponds to a small temperature part of the second relativistic regime found for the Alfvenic wave.
However, four temperature regimes for the fast magneto-sound.
So, the second relativistic regime found for the Alfvenic wave splits for two areas.

Further increase of temperature up to $\beta=0.002$
shows further change of
the low-frequency part of the fast extraordinary wave
and the small frequency second branch.
It corresponds to the small relativistic temperature effects on ions.
So, the effective cyclotron frequency $\Omega_{i}\sqrt{(1-\frac{5U_{pe,i}^{2}}{c^{2}})}$ is not changed in compare with $\Omega_{i}$.
The small frequency second branch shows further increase of the refractive index.
Moreover, its behavior is changed near the ion cyclotron frequency.
Tendency $n^{2}\rightarrow -\infty$ is changed to $n^{2}\rightarrow +\infty$.
The refractive index of the fast extraordinary wave is also changed.
In this regime, the minimal frequency of the fast extraordinary wave $\omega_{min}>\Omega_{i}$.
Fig. 8 shows $\omega_{min}/\omega_{Le}=0.00017$.
At $\omega/\omega_{Le}\in(1.7\times10^{-4}, 1.85\times10^{-4})$ the refractive index of the fast extraordinary wave is smaller
then frequency of the fast magneto-sound wave existing in the nonrelativistic limit.
At larger frequencies $\omega/\omega_{Le}\in(1.85\times10^{-4}, 0.2)$
the refractive index of the fast extraordinary wave is larger then frequency of the fast magneto-sound wave.

Regime of relativistic ions with the temperature comparable with the rest energy of ion is presented in Fig. 9.
Such large temperature leads to the negative square of the effective cyclotron frequency $\Omega_{i}^{2}(1-\frac{5U_{pi}^{2}}{c^{2}})$.
Hence, there is no pole for this frequency.
Therefore, the small frequency second branch existing below this frequency does not exist in this regime.
We find the low-frequency part of the fast extraordinary wave only.
The fast extraordinary wave exist at $\omega>\hat{\omega}$.
At $\beta=0.0005$ we have $\hat{\omega}/\omega_{Le}=0.00015$.
At $\omega/\omega_{Le}>0.0002$
the refractive index of the fast extraordinary wave is above corresponding values for the nonrelativistic fast magneto-sound (see the upper Fig. 9).
The lower Fig. 9 shows that such behavior goes up to $\omega/\omega_{Le}=0.2$.
At larger frequencies $\omega/\omega_{Le}\in(0.2,0.5)$ the refractive index of the fast extraordinary wave is smaller
in compare with the nonrelativistic fast magneto-sound.

\section{Conclusion}

A dramatic change of the spectrum of the Alfvenic waves and fast magneto-sound has been found for the relativistically hot plasmas.
First, these changes are crucial in compare with the nonrelativistic regime.
Next, these changes are different for various relativistic regime
which have been considered in the paper.
More exactly, three relativistic regimes have been considered for the relativistically hot plasmas.
These regimes corresponds to the different temperature intervals for the Alfvenic waves.
Four regimes are found for different temperature intervals for the fast magneto-sound.
Changes of spectra have been found.
Modifications of areas of existence of waves have been obtained.

\section{Acknowledgements}

Work is supported by the Russian Foundation for Basic Research (grant no. 20-02-00476).
This paper has been supported by the RUDN University Strategic Academic Leadership Program.

\section{DATA AVAILABILITY}

Data sharing is not applicable to this article as no new data were
created or analyzed in this study, which is a purely theoretical one.

\appendix

\section{Appendix A: Numerical estimation of the relativistic-thermal parameters for electrons}

For the electrons we have $\chi_{e}=1$,
so $f_{1,2,3}(\chi_{e}\beta)=f_{1,2,3}(\beta)$.

Seven temperature regimes are chosen for the electrons
$T_{e1}=0.1 m_{e}c^{2}$, $T_{e2}=m_{e}c^{2}$, $T_{e3}=10m_{e}c^{2}$, $T_{e4}=2\times 10^{3} m_{e}c^{2}$,
$T_{e5}=0.33m_{e}c^{2}$, $T_{e6}=0.5m_{e}c^{2}$, and $T_{e7}=0.5 \times 10^{3} m_{e}c^{2}$.
It gives the following values of the dimensionless reverse temperature $\beta=m_{e}c^{2}/T_{e}$ and corresponding values of other parameters.

For the relatively small relativistic temperature $\beta_{1}=10$,
we calculate $K_{1}/K_{2}=0.91$, $U_{t}^{2}/c^{2}=0.07$,
$U_{p}^{2}/c^{2}=0.08$, and $U_{M}^{4}/c^{4}=0.02$,
where
$K_{1}(10)=2\times 10^{-5}$, $K_{2}(10)=2.2\times10^{-5}$, $f_{1}(10)=5\times 10^{-7}$,
$f_{2}(10)=4.2\times10^{-7}$, $f_{3}(10)=1.7\times 10^{-7}$.

For $\beta_{5}=3$,
we get $K_{1}/K_{2}=0.65$,
$U_{t}^{2}/c^{2}=\beta f_{2}/(3 K_{2})=0.11$,
$U_{p}^{2}/c^{2}=\beta f_{1}/(3 K_{2})=0.18$,
and $U_{M}^{4}/c^{4}=\beta f_{3}/(5 K_{2})=0.07$,
where
$K_{1}(3)=0.0402$,
$K_{2}(3)=0.062$,
$f_{1}(3)=0.011$,
$f_{2}(3)=0.0066$,
$f_{3}(3)=0.0071$.

For $\beta_{6}=2$,
we get $K_{1}/K_{2}=0.55$,
$U_{t}^{2}/c^{2}=\beta f_{2}/(3 K_{2})=0.11$,
$U_{p}^{2}/c^{2}=\beta f_{1}/(3 K_{2})=0.22$,
and $U_{M}^{4}/c^{4}=\beta f_{3}/(5 K_{2})=0.1$,
where
$K_{1}(2)=0.14$,
$K_{2}(2)=0.25$,
$f_{1}(2)=0.084$,
$f_{2}(2)=0.042$,
$f_{3}(2)=0.062$.

For $\beta_{2}=1$, we get $K_{1}/K_{2}=0.38$, $U_{t}^{2}/c^{2}=\beta f_{2}/(3 K_{2})=0.1$,
$U_{p}^{2}/c^{2}=\beta f_{1}/(3 K_{2})=0.28$, and $U_{M}^{4}/c^{4}=\beta f_{3}/(5 K_{2})=0.15$,
where $K_{1}(1)=0.6$, $K_{2}(1)=1.6$, $f_{1}(1)=1.35$, $f_{2}(1)=0.46$, $f_{3}(1)=1.17$.

For $\beta_{3}=0.1$,
we have $K_{1}/K_{2}=0.05$,
$U_{t}^{2}/c^{2}=\beta f_{2}/(3 K_{2})=0.02$,
$U_{p}^{2}/c^{2}=0.33$,
and $U_{M}^{4}/c^{4}=0.2$,
where $K_{1}(0.1)=10$,
$K_{2}(0.1)=200$,
$f_{1}(0.1)=2\times 10^{3}$,
$f_{2}(0.1)=100$,
$f_{3}(0.1)=2\times 10^{3}$.

For $\beta_{7}=2\times 10^{-3}$,
we find
$K_{1}/K_{2}=1 \times 10^{-3}$,
$U_{t}^{2}/c^{2}=\beta f_{2}/(3 K_{2})=3.3 \times 10^{-4}$,
$U_{p}^{2}/c^{2}=\beta f_{1}/(3 K_{2})=0.33$,
and
$U_{M}^{4}/c^{4}=\beta f_{3}/(5 K_{2})=0.2$,
where
$K_{1}(2\times 10^{-3})=5 \times 10^{2}$,
$K_{2}(2\times 10^{-3})=5 \times 10^{5}$,
$f_{1}(2\times 10^{-3})=2.5\times 10^{8}$,
$f_{2}(2\times 10^{-3})=2.5\times 10^{5}$,
$f_{3}(2\times 10^{-3})=2.5\times 10^{8}$.

For $\beta_{4}=0.5\times 10^{-3}$,
we find
$K_{1}/K_{2}=0.25 \times 10^{-3}$,
$U_{t}^{2}/c^{2}=\beta f_{2}/(3 K_{2})=8.3 \times 10^{-5}$,
$U_{p}^{2}/c^{2}=\beta f_{1}/(3 K_{2})=0.33$,
and
$U_{M}^{4}/c^{4}=\beta f_{3}/(5 K_{2})=0.2$,
where
$K_{1}(5\times 10^{-4})=2\times 10^{3}$,
$K_{2}(5\times 10^{-4})=8\times 10^{6}$,
$f_{1}(5\times 10^{-4})=1.6\times 10^{10}$,
$f_{2}(5\times 10^{-4})=4\times 10^{6}$,
$f_{3}(5\times 10^{-4})=1.6\times 10^{10}$.

\section{Appendix B: Numerical estimation of the relativistic-thermal parameters for ions}

If we consider the isothermal plasmas $T_{i}=T_{e}$
we would also have seven temperature regimes.

However, functions $f_{1,2,3}(\chi_{i}\beta)$
depend on $\chi_{i}$
which should be calculated in each regime.
Parameter $\chi_{i}=(m_{i}/m_{e})\cdot(T_{e}/T_{i})$ simplifies in the isothermal regime.
$\chi_{i,I}=(m_{i}/m_{e})$,
where symbol $I$ refers to the isothermal regime.
Here, we consider the electron-proton or hydrogen plasmas.
Hence, parameter $\chi_{i,I}=(m_{i}/m_{e})=1.84\times 10^{3}$ is fixed.
The argument of functions $f_{1,2,3}$ has the following values:
$\chi_{i,I}\beta_{1}=1.84\times 10^{4}$,
$\chi_{i,I}\beta_{2}=1.84\times 10^{3}$,
$\chi_{i,I}\beta_{3}=1.84\times 10^{2}$,
and $\chi_{i,I}\beta_{4}=0.92 $.

We start from the small temperature regime
$\chi_{i,I}\beta_{1}=1.84\times 10^{4}$,
we calculate $K_{1}/K_{2}=0.99992$,
$U_{t}^{2}/c^{2}=0$,
$U_{p}^{2}/c^{2}=\chi_{i,I}\beta f_{1}(\chi_{i,I}\beta)/(3 K_{2}(\chi_{i,I}\beta))=0$,
and $U_{M}^{4}/c^{4}=0$,
where
$f_{1}(1.84\times 10^{4})=0$,
$f_{2}(1.84\times 10^{4})=0$,
$f_{3}(1.84\times 10^{4})=0$.

For $\chi_{i,I}\beta_{2}=1.84\times 10^{3}$,
we get
$K_{1}/K_{2}=0.9992$,
$U_{t}^{2}/c^{2}=0$,
$U_{p}^{2}/c^{2}=0$,
and $U_{M}^{4}/c^{4}=0$,
where
$f_{1}(1.84\times 10^{3})=0$,
$f_{2}(1.84\times 10^{3})=0$,

For $\chi_{i,I}\beta_{6}=3.68\times 10^{3}$,
we get
$K_{1}/K_{2}=0.9996$,
$U_{t}^{2}/c^{2}=0$,
$U_{p}^{2}/c^{2}=0$,
and $U_{M}^{4}/c^{4}=0$,
where
$f_{1}(1.84\times 10^{3})=0$,
$f_{2}(1.84\times 10^{3})=0$,
$f_{3}(1.84\times 10^{3})=0$.

For $\chi_{i,I}\beta_{5}=5.52\times 10^{3}$,
we get
$K_{1}/K_{2}=0.9997$,
$U_{t}^{2}/c^{2}=0$,
$U_{p}^{2}/c^{2}=0$,
and $U_{M}^{4}/c^{4}=0$,
where
$f_{1}(1.84\times 10^{3})=0$,
$f_{2}(1.84\times 10^{3})=0$,
$f_{3}(1.84\times 10^{3})=0$.

For $\chi_{i,I}\beta_{3}=1.84\times 10^{2}$,
we have
$K_{1}/K_{2}=0.992$,
$U_{t}^{2}/c^{2}=5.3\times 10^{-3}$,
$U_{p}^{2}/c^{2}=5.3\times 10^{-3}$,
and $U_{M}^{4}/c^{4}=8.5\times 10^{-5}$,
where
$K_{1}(1.84\times 10^{2})=1.14 \times 10^{-81}$,
$K_{2}(1.84\times 10^{2})=1.2\times 10^{-81}$,
$f_{1}(1.84\times 10^{2})=0$, 
$f_{2}(1.84\times 10^{2})=0$, 
$f_{3}(1.84\times 10^{2})=0$. 

For $\chi_{i,I}\beta_{4}=0.92$ corresponding to temperature close to the rest energy of the proton,
we have
$K_{1}/K_{2}=0.35$,
$U_{t}^{2}/c^{2}=0.09$,
$U_{p}^{2}/c^{2}=0.28$,
and $U_{M}^{4}/c^{4}=0.15$,
where
$K_{1}(0.92)=0.69$,
$K_{2}(0.92)=1.98$,
$f_{1}(0.92)=1.82$,
$f_{2}(0.92)=0.58$,
$f_{3}(0.92)=1.6$.

For $\chi_{i,I}\beta_{7}=3.68$
corresponding to temperature close to the rest energy of the proton,
we have
$K_{1}/K_{2}=0.7$,
$U_{t}^{2}/c^{2}=0.1$,
$U_{p}^{2}/c^{2}=0.16$,
and $U_{M}^{4}/c^{4}=0.06$,
where
$K_{1}(3.68)=1.8 \times 10^{-2}$,
$K_{2}(3.68)=2.6 \times 10^{-2}$,
$f_{1}(3.68)=3.4 \times 10^{-3}$,
$f_{2}(3.68)=2.2 \times 10^{-3} $,
$f_{3}(3.68)=2 \times 10^{-3}$.

\end{document}